\documentclass{article}
\usepackage[a4paper, total={5in, 8in}]{geometry}

\usepackage{hyperref}

\usepackage[english]{babel}
\usepackage[utf8]{inputenc}
\usepackage[T1]{fontenc}
\usepackage{amsthm}
\usepackage{thmtools}
\usepackage{thm-restate}
\usepackage{dsfont}

\declaretheorem{definition}
\declaretheorem{theorem}

\usepackage{amsmath}
\usepackage{amsfonts}
\usepackage{cleveref}
\usepackage{xspace}
\usepackage{color, xcolor}
\usepackage{todonotes}
\usepackage{listings}
\usepackage{multicol}
\usepackage{booktabs, multirow}

\usepackage{tikz}

\usepackage{listings}
\usepackage{lstcoq}
\usepackage{setspace}

\usepackage[noend]{algpseudocode} %
\usepackage{algorithm}

\lstnewenvironment{coq}{\lstset{language=Coq,basicstyle=\small}}{}

\newtheorem{coro}{Corollary}
\newtheorem{goodsit}{Good Situation}

\newcommand{\lb}{\ensuremath{\mathtt{LowerBoundBS}}\xspace}

\newcommand{\lbcoq}{\ensuremath{\mathtt{LowerBoundCoq}}\xspace}
\newcommand{\oi}{\ensuremath{\mathtt{OnlineInfeasible}}\xspace}
\newcommand{\loads}{\ensuremath{\mathit{Lo}}\xspace}
\newcommand{\items}{\ensuremath{\mathcal{I}}\xspace}
\newcommand{\codeloads}{\ensuremath{\mathtt{Lo}}}
\newcommand{\codeitems}{\ensuremath{\mathtt{I}}}
\newcommand{\nil}{\ensuremath{\mathit{nil}}\xspace}

\newcommand{\myproblem}[1]{\textsc{#1}}
\newcommand{\binstretching}{\myproblem{Online Bin Stretching}\xspace}

\newcommand\Game{{\mathrm{BSG}}}
\newcommand\algoplayer{\textsc{Algorithm}\xspace}
\newcommand\adveplayer{\textsc{Adversary}\xspace}
\newcommand\algo{\algoplayer}
\newcommand\adversary{\adveplayer}
\newcommand\evaladv{\textsc{EvaluateAdversary}\xspace}
\newcommand\evalalg{\textsc{EvaluateAlgorithm}\xspace}

\newcommand\Sequential{\textsc{Sequential}\xspace}

\newcommand\DynprogMax{\textsc{DynprogMax}\xspace}
\newcommand\Query{\textsc{Query}\xspace}
\newcommand\MaxFeas{\textsc{MaxFeas}\xspace}
\newcommand\bfd{\textsc{Best Fit Decreasing}\xspace}
\newcommand\obf{\textsc{Online Best Fit}\xspace}
\newlength{\elseskiplen}
\newcommand\elseskip{\hskip\elseskip}
\newcommand\eps\varepsilon

\newcommand{\bigO}{\ensuremath{\mathcal{O}}}

\makeatletter
\newcommand*{\defeq}{\mathrel{\rlap{%
                     \raisebox{0.3ex}{$\m@th\cdot$}}%
                     \raisebox{-0.3ex}{$\m@th\cdot$}}%
                     =}
\makeatother

\newcommand\parperiod{}

\title{Discovering and Certifying Lower Bounds for the \\Online Bin Stretching Problem}
\author{
    Martin B\"ohm, \textsuperscript{\hskip-1.5mm\rm 1}
    Bertrand Simon \textsuperscript{\hskip-1.5mm\rm 2}
}
\date{
\small{

  \textsuperscript{\rm 2} Institute of Computer Science, University of Wrocław, Poland\\
  \textsuperscript{\rm 1} IN2P3 Computing Center / CNRS, Lyon - Villeurbanne, France\\
    boehm@cs.uni.wroc.pl, bertrand.simon@cc.in2p3.fr
}}

\begin{document}
	
	\maketitle

\begin{abstract}

There are several problems in the theory of online computation where
tight lower bounds on the competitive ratio are unknown and expected
to be difficult to describe in a short form. A good example is the
\myproblem{Online Bin Stretching} problem, in which the task is to pack
the incoming items online into bins while minimizing the load of the
largest bin. Additionally, the optimal load of the entire instance is
known in advance.

The contribution of this paper is twofold. We use the Coq proof assistant to
formalize the \myproblem{Online Bin Stretching} problem and provide a program
certifying lower bounds of this problem. Because of the size of the
certificates, previously claimed lower bounds were never formally proven. To the
best of our knowledge, this is the first use of a formal verification toolkit to
certify a lower bound for an online problem.

We also provide the first
non-trivial lower bounds for \myproblem{Online Bin Stretching} with 6,
7 and 8 bins, and increase the best known lower bound for 3 bins. We
describe in detail the algorithmic improvements which were necessary
for the discovery of the new lower bounds, which are several orders of
magnitude more complex. 
\end{abstract}

\newpage

\section{Introduction}
\label{sec:intro}

The problem \binstretching has been introduced by Azar and
Regev~\cite{azar2001binstretch} as a semi-online generalization of the
\myproblem{Online Bin Packing} problem. Specifically, the task consists
of packing various-size elements (items) arriving in an online fashion
into $m$ different bins. The problem belongs to the category of
semi-online problems as there is a guarantee (known beforehand) that
all the input items can be packed into $m$ bins of a given size
$g$. The objective is to minimize the load  of the largest bin. The
performance measure (here named the \emph{stretching factor}) of an
online algorithm is the maximum for all inputs of the load of the
largest bin divided by $g$.

Note that this setting is equivalent to the classical scheduling
problem \textsc{Online Makespan Minimization} where the optimal makespan
of the instance is known in advance to the algorithm.

\paragraph*{Lower bounds for \binstretching}

\ \smallskip

The earliest lower bound for \binstretching is due to Kellerer
et. al.~\cite{kellererkotov97}, dating even before the introductory
paper of Azar and Regev~\cite{azar2001binstretch}. Kellerer
et. al.~\cite{kellererkotov97} show a lower bound of $4/3$ for the
case of $2$ bins, while Azar and Regev~\cite{azar2001binstretch} extend
it to an arbitrary number of bins. In the special case of 2 bins, it is
known that this lower bound is tight, as there is an algorithm with
stretching factor $4/3$~\cite{kellererkotov97}.

On the algorithmic front, the earliest algorithm for any number of
bins~\cite{azar2001binstretch} achieves a stretching factor of
$1.625$. More efficient algorithms have since been proposed, and the
current best algorithms designed by Böhm et. al.~\cite{bohm2014algo}
have a stretching factor of $1.5$ for any number of bins and
$11/8=1.375$ for exactly 3 bins.

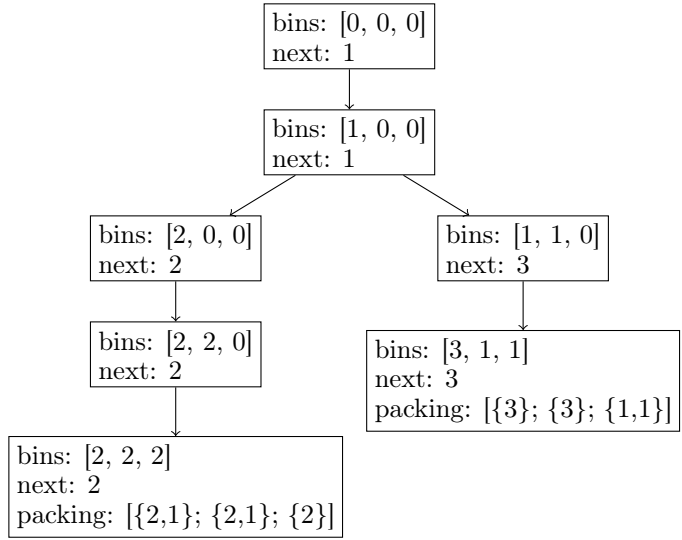
\begin{figure}[tb]
	\centering
	\newcommand{\nnode}[4]{bins: [#1, #2, #3]\\[-1pt] next: #4}
	\newcommand{\nnodep}[5]{bins: [#1, #2, #3]\\[-1pt] next: #4\\[-1pt] packing: #5}
	\begin{tikzpicture}[every node/.style={rectangle, draw, align=left, inner sep = 3pt, node distance = 40pt}]
	\node  (1) {\nnode0001};
	\node[below of = 1] (2) {\nnode1001};
	\node[below of = 2, xshift = -65pt] (3a) {\nnode2002};
	\node[below of = 2, xshift = 65pt] (3b) {\nnode1103};
	\node[below of = 3a] (4a) {\nnode2202};
	\node[below of = 3b, yshift=-10pt] (4b) {\nnodep3113{[\{3\}; \{3\}; \{1,1\}]}};
	\node[below of = 4a, yshift=-10pt] (5a) {\nnodep2222{[\{2,1\}; \{2,1\}; \{2\}]}};
	\draw[->] (1) edge (2) (2) edge (3a) (3a) edge (4a) (4a) edge (5a)  (2) edge (3b) (3b) edge (4b);
	\end{tikzpicture}
	\caption{Tree describing the lower bound of $4/3$ for 3 bins of size 3. In each node, the first list represents the current load of each bin and the second number represents the next item to appear in the online instance. At the leaves, a packing of the relevant items in bins of size 3 is provided.}
	\label{fig:lb43}
\end{figure}

The original lower bound of $4/3$ for any number of bins $m$ is
depicted in \Cref{fig:lb43}. Each node of the tree corresponds to a
state of the online process: the online algorithm has packed the
current items in the bins, and the next item of the instance is
provided. Each child of a node represents a possible choice for the
online algorithm in which all bin sizes are less than $4$. Every leaf
node contains a proof of existence of a packing that fits all items
into $m$ bins of capacity $3$.

Since the earliest publication of the lower bound above in 1997 (for
two bins, \cite{kellererkotov97}), significant effort has been spent
by several research groups in order to discover a new lower bound with
a better ratio. Despite those efforts, no better lower bound is known
for general $m$.

Positive progress has been made for cases with small fixed values of
$m$. Gabay, Brauner and Kotov \cite{gabay2013} present a lower
bound of $19/14\approx1.357$ for $m=3$ using an extensive computer
search, essentially an implementation of the minimax algorithm in this
setting.

As the binary result (true or false) of such a computer program should
not be blindly trusted, as it is prone to human error, they produced a
decision tree, similar to \Cref{fig:lb43} in order to prove the
result. Printing out this tree required a 6-page appendix and it can
therefore still be verified by a human, but such a task is quite
tedious. This lower bound has been independently generalized to $m=4$
machines by Gabay et. al.~\cite{gabay2017improved} and Böhm et.
al~\cite{bohm2017LB}, and the strategy was already too large to be
printed on paper. Subsequent research by Böhm et. al.~\cite{bohm2017LB}
leads to the current state described in ~\Cref{tbl:LB}. The
first contribution of this paper is to extend these results as
presented in \Cref{tbl:LBnew}. Specifically, the lower bound of
$19/14$, which was already established for $m\in\{2,3,4,5\}$, is now
established for the settings $m\in\{6,7,8\}$. For $m=3$, which is the
only setting for which a lower bound larger than $19/14$ is known, we
have also improved it from $45/33$ to $112/82$ (recall that the best upper bound here equals $11/8=112.75/82$). It should be noted
that the size of the trees involved has dramatically increased, going
from a few thousands of nodes to billions of nodes. This is the
consequence of several major improvements in the computer program,
which were previously described in the PhD thesis of one of the
authors~\cite{bohm2018phd}, and which we detail in this paper.

\paragraph{Certified algorithms}
\ \smallskip

Due to the enormous
increase in the size of the strategy output, the aforementioned
researchers had to resort to a separate program, which can be called a
\emph{checker}, in order to verify the validity of the tree.
Therefore, the lower bounds proved so far depend on the correctness of
this checker program. It should be noted that the trees are not
actually stored as explicitly, but rather use a DAG structure in order
to avoid duplicate subtrees.

\begin{table*}[tb]
	\begin{tabular}{ccccccccc} \toprule
		Value of $m$  & $2$ & $3$ & $4$ & $5$ & any \\ \midrule
		{Upper bound} & 4/3 & 1.375 & 1.5 & 1.5 & 1.5 \\
		\multirow{2}{*}{Lower bound} & 4/3 & 45/33 & 19/14 & 19/14 & 4/3 \\
		 & 1.333 & 1.3636 & 1.357 & 1.357 & 1.333\\
		Tree nodes & 5 & 5080 & 433 & 3908 &  \\		
		\bottomrule
	\end{tabular}
	\caption{Previously known lower bounds and number of nodes in the tree describing them. }
	\label{tbl:LB}
\end{table*}

\begin{table*}[tb]
	\begin{tabular}{ccccccccc} \toprule
		Value of $m$  & $2$ & $3$ & $4$ & $5$ & $6$ & $7$ & $8$ \\ \midrule
		\multirow{2}{*}{Lower bound} & 4/3 & 112/82 & 19/14 & 19/14 & 19/14 & 19/14 & 19/14 \\
		& 1.333 & 1.3658 & 1.357 & 1.357 & 1.357 & 1.357 & 1.357  \\
		Tree nodes & 5 & $186$k & 433 & 3908 & $3.8$M & $231$M & $2.5$G\\		
		\bottomrule
	\end{tabular}
	\caption{Current known lower bounds and number of nodes in the tree describing them.}
		\label{tbl:LBnew}
	\end{table*}

This method of computing lower bounds falls into the definition of
certifying algorithms, which were introduced by Blum and Kannan in~\cite{blum1995certif}. Such an
algorithm can be defined as providing a \emph{certificate}, or a
\emph{witness} in addition to the classic output: given an input $x$,
it computes the output $y$ and provides a witness $w$. The certifying algorithm
is accompanied by a checker program, which is typically much simpler,
and which can verify, given $x$, $y$, and $w$, that $y$ is a valid
solution. In our context, the witness corresponds to the tree
describing the strategy. Such a strategy has for instance been adopted
in the algorithmic library LEDA~\cite{mehlhorn1997leda} concerning the maximum
cardinality matching problem on graphs. The remaining drawback of this
approach is that the checker program still has to be correct in order
to trust the solution $y$. While the program was arguably simple, 
Alkassar et. al.~\cite{alkassar2014certif} used the automatic verifier VCC and the
interactive theorem prover Isabelle~\cite{nipkow2002isabelle} in order to build a formal
proof of the correctness of the checker program. Surprisingly, there
was a bug in the checker, which could make it accept a wrong solution
for some ill-formed witness. For a complete survey on the domain of
certifying algorithms, we refer the reader to~\cite{mcconnell2011certifsurvey}.
Subsequent works on checker verification can be found in~\cite{trustworthyGraph,noschinski2016formalizing,ShortestPath-AFP}.

The lesson that one can learn from this example is that it is arguably dangerous
to base a result on the output of a non-trivial program, even if this program
seems simple, such as the checker of the online bin stretching lower bounds. The
second contribution of this paper is therefore to provide a certified checker.
Specifically, we use the proof assistant Coq~\cite{barras1997coq} to formalize
the \binstretching problem. Then, we build a checker in the Gallina language
used by Coq. We prove that if this checker returns \verb|true| given a strategy
tree, then the corresponding lower bound is valid. Finally, we run the program
on the existing trees in order to certify their validity. This program has
been developed to be easy to be reused in the future in order to certify new
lower bounds. To the best of our knowledge, this is the first time that a proof
assistant software is used to certify such a lower bound found by computer
search. We hope that this contribution will help to establish the standard
of formally verifying results that cannot be properly checked by a human.

It should be noted that we do not provide any certified result if the
computer search procedure does not find a lower bound. As the item
sizes are constrained in the computer search, there may exist a
lower bound requiring other item sizes.

The rest of the paper is organized as follows. In \Cref{sec:def}, we
formally define the problem. In \Cref{sec:search}, we describe in detail the
program that we used to improve the best known lower bounds via
computer search. In \Cref{sec:coqform}, we propose a formalization of
the \emph{lower bound} property in Coq, prove that this property
matches our definition, and detail the results obtained on the best
known lower bounds. Note that we do not detail here the Coq proofs nor the
checker, as they are not necessary to prove the result. The complete
code is available online at~\cite{GithubSearch} and~\cite{GithubCoq}.

\section{Bin stretching as a two-player game}
\label{sec:def}
In this section, we formally define \binstretching with integer-sized
items as an equivalent two-player game with the two players named
\algoplayer and \adveplayer.

The bin stretching game $(\Game)$ will be parameterized by three positive
integers $m,t$ and $g$. Before proceeding formally, we wish to note
that $m$ stands for the number of bins \emph{(machines)} in the
instance, $t$ stands for the \emph{target} of the \algoplayer and $g$
corresponds to the \emph{guarantee} that the \adveplayer must satisfy.

\begin{definition}\label{dfn:game}
The bin stretching game $\Game(m,t,g)$ is a two player game defined as follows:
During each round (indexed by $i$), the player \adveplayer chooses a
positive integer $e_i$, corresponding to the size of the next item
of the input sequence. After that, the player \algoplayer chooses a
bin index $b_i$ between $0$ and $m-1$, into which he packs this
item. The player \adveplayer wins if and only if there exists a
round $j$ such that:

\begin{enumerate}
\item (Hitting the target.) \algoplayer loads a bin to capacity $t$, i.e., there exists a bin index $b$ such that
  $\sum_{i\leq j}  (\mathds{1}_{b_i=b}\cdot {e_i})\geq t$, where $\mathds{1}_{b_i=b}$ equals $1$ if $b_i=b$ and 0 otherwise.
\item (The guarantee.) There exists a packing of the items $\{e_i,~ i\leq j\}$ into $m$ bins with capacity at most $g$.
\end{enumerate}
\end{definition}

Note that after some amount of rounds (at most during the $gm$-th
round), \adveplayer cannot win as any subsequent packing will have a
load of at least $g+1$. Thus, whenever the player \adveplayer is unable
to present any item, we note the game state as winning for the player
\algoplayer.

{We will often refer to a specific state of the game in progress,
and we wish to describe two possible representations of such a state.}

\begin{definition}\label{dfn:notation}
A \emph{bin configuration} is a state of the game before the player \adveplayer
makes a move (presents an item). Such a configuration can be represented as a pair
$(\items,\loads)$, where $\loads$ is an $m$-tuple of \emph{loads} of the bins and
$\items$ is a list of items such that these items can be packed into $m$ bins
forming exactly the $m$-tuple $\loads$.

We also define the \emph{extended representation} of a bin
configuration as the $m$-tuple $[E_1, E_2, \ldots, E_m]$, where each
$E_i$ is a list of items which are currently packed into bin $i$.
\end{definition}

For example, suppose $m=3$ and the following items were presented
by the player \adversary so far: $[1,1,2,3,4]$. Then, one possible bin
configuration might be $([1,1,2,3,4], [5,4,2])$ with the extended
representation being $[[4,1],[3,1],[2]]$.

A careful reader will observe there can be another extended
representation of the same bin configuration, namely
$[[3,2],[4],[1,1]]$. However, it is true that, from the point of view of
both \algoplayer and \adveplayer, the game state is the same -- the
loads are the same and the sequence of items is also. There is nothing
in the second representation that either player can use to their
benefit compared to the first representation. Thus, it is correct to
treat them as variants of a single bin configuration.

The property of existence of a winning strategy for player \adveplayer
is clearly the main goal of our efforts. We define it formally, so
that we can refer to it later, during our verification efforts:

\begin{restatable}{definition}{propertylb}
\label{dfn:propertylb}
If the player \adveplayer has a winning strategy, we say that $\Game(m,t,g)$
satisfies the property \lb. This implies that no online algorithm can
solve the \binstretching problem with a stretching factor smaller than
$t/g$.

If the player \adveplayer has a winning strategy for the extended game
with a starting bin configuration $(\items,\loads)$, we say that
$\Game(m,t,g)$ satisfies the property $\lb(\items,\loads)$.

\end{restatable}

If $\loads$ is composed only of zeros, we then have by
definition: $$\lb(\nil,\loads) \Longrightarrow \lb.$$

\subsection{Notation}\label{sec:notation}

\paragraph{Sizes\parperiod} Most of the time, it is convenient
for us to use the variable of the item $e$ interchangeably as its
size. When it is useful to distinguish these two concepts, we also use
the function $s(e)$ for the size of item $e$. Extending the notation,
we also use $s(A)$ for the current load of bin $A$ (sum of the items
packed within) and $s(\mathcal{T})$ for the total load of a set, list,
or tuple of bins $\mathcal{T}$.

\paragraph{Set operations on bin configurations\parperiod} The
bin configuration $(\items,\loads)$ consists of the $m$-tuple of loads
$\loads$ and list (multiset) of items $\items$. We use set-theoretical
notation to refer to individual bins within $\loads$; for example, we
may choose an arbitrary bin $A \in \loads$ or consider $\loads$ without
the largest bin $B_1$, which we would denote as $\loads \setminus B_1$.

\section{Computing new lower bounds via computer search}
\label{sec:search}

Our implemented algorithm is a parallel, multi-computer implementation of the
classical \textsc{minimax} game search algorithm. We now describe a pseudocode
of its sequential version. The main procedure of the \textsc{minimax} algorithm
is the procedure $\Sequential$ as stated below, which recursively calls the
evaluation subroutines $\evaladv$ and $\evalalg$. The peculiarities of our
algorithm (caching, pruning, parallelization) are described in the following
sections.

One of the differences between our algorithm and the algorithm of
Gabay et. al. \cite{gabay2017improved} is that {our bin stretching
game from \Cref{dfn:game} contains no concept of payoff, and thus} our
algorithm makes no use of alpha-beta pruning -- indeed, as either
\algo or \adversary has a winning strategy from each bin
configuration, there is no need to use this type of pruning.

\begin{algorithm}
\caption{Procedure $\evaladv$}
Input is a bin configuration $C = (\items,\loads)$.
\begin{algorithmic}[1]
\State \algorithmicif\ the configuration is cached (Section~\ref{sec:4:caching}), \Return the value found in cache.
\State Create a list $L$ of items which can be sent as the next step of the player \adversary (Section \ref{subsec:test}).
\For{every item size $i$ in the list $L$}
\State Recurse by running $\evalalg(C,i)$.
\State \algorithmicif\ $\evalalg(C,i)$ returns $0$ (the configuration is winning for player \adversary), stop the cycle and \Return $0$.
\EndFor
\State \algorithmicif\ the evaluation reaches this step, store the configuration in the cache and \Return $1$ (player \algo wins).

\end{algorithmic}
\end{algorithm}

\begin{algorithm}
\caption{Procedure $\evalalg$}
Input is a bin configuration $C = (\items,\loads)$ and item $i$.
\begin{algorithmic}[1]
\State Prune the tree using known algorithms (Section \ref{sec:4:gs}).
\For{any one of the $m$ bins}
\If{$i$ can be packed into the bin so that its load is at most $t-1$}
\State Create a configuration $C'$ that corresponds to this packing.
\State Run $\evaladv(C')$.
\State \algorithmicif\ $\evaladv(C')$ returns 1, \Return 1 as well.
\EndIf
\EndFor
\State \algorithmicif\ we reach this step, no placement of $i$ results in victory of \algo; \Return 0.
\end{algorithmic}
\end{algorithm}

\begin{algorithm}
\caption{Procedure $\Sequential$}
\noindent Input is a bin configuration $C = (\items,\loads)$. 
\begin{algorithmic}[1]
\State Fix parameters $m,t,g$.
\State Run $\evaladv(C)$.
\If{$\evaladv(C)$ returns $0$}
\State \Return success (a lower bound exists).
\Else
\State \Return failure.
\EndIf
\end{algorithmic}
\end{algorithm}
\subsection{Verifying the offline optimum guarantee}\label{subsec:test}

When we evaluate a turn of the \adversary, we need to create the list $L =
\{0,1,\ldots,y\} \subseteq \{0,1,\ldots,g\}$ of items that \adversary can
actually send while satisfying the \binstretching guarantee. In other words, we
compute the value $y$ representing the maximum item size that the adversary can
send while satisfying the guarantee. We do this operation inside the procedure $\MaxFeas$, which we describe in
\Cref{sec:4:dynprogbounds}.

\subsubsection{Procedure $\MaxFeas$}\label{sec:4:dynprogbounds}

If we wish to directly compute the \textit{maximum feasible} value $y$
which can be sent from the configuration $(\items, \loads)$ where
$|\items| = n$, we can do so by calling the dynamic program
$\DynprogMax$ (\Cref{sec:4:dynprogmax}). The complexity of
$\DynprogMax$ is $O(n \cdot g^m)$ in the worst-case scenario (if we
ignore potential slowdowns via hashing).

This is polynomial when $m$ is a constant, but already for $m=3$ and
especially when $4 \le m \le 8$ such a call per game state
becomes prohibitively expensive. Therefore, we first compute estimates
$LB$ and $UB$ on the value $y$ such that $LB \le y \le UB$. Ideally,
our faster estimates can get to the ideal value directly, making the
dynamic programming call unnecessary.

We first initialize the upper bound from previous computations.  The
upper bound will be set as $UB \defeq \min(y', m\cdot g - V)$, where
$y'$ is the maximum feasible value that was computed in the previous
vertex of \adversary's turn, and $V$ is the total size of all items in
the instance. The second term $m\cdot g - V$ is therefore the sum of
all items that can yet arrive in this instance.

\paragraph{Online Best Fit\parperiod} To find the first lower bound on $y$ quickly, we
employ an online bin packing algorithm \obf. This algorithm maintains
a packing of items $\items$ to $m$ bins of size $g$ during the
evaluation of the algorithm $\Sequential$, packing each item as it is
selected by the player \adversary. The algorithm \obf packs each item
$i$ into the most-loaded bin where the item fits.

Once the algorithm $\Sequential$ selects a different item $i'$ and
evaluates a different branch of the game tree, \obf removes $i$ from
its bin and inserts $i'$ to the most-loaded bin where $i'$ fits.

As \obf maintains just one packing, which may not be optimal, it can
happen that \obf is unable to pack the next item $i$ even though $i$
is a feasible item. In that case, we mark the packing as inconsistent
and do not use the lower bound from \obf until its online packing
becomes feasible again.

If the packing maintained by \obf is still feasible, we return as the
lower bound value $LB$ the amount of unused space on the least-loaded
bin.

The main advantage of \obf is that it takes at most $O(m)$ time per
each step, and especially for the earlier stages of the evaluation its
returned value can match the value of $y$.

\paragraph{Checking the cache\parperiod} Next, if a gap still remains between
$LB$ and $UB$, we try to tighten it by calling a procedure \Query
which queries the cache of feasible and infeasible item multisets.
The procedure has a ternary answer -- either an item multiset $\items
\cup \{j\}$ was previously computed to be feasible, or it was computed
to be infeasible, or this item set is not present in the cache at all.%

We update $LB$ to be the largest value which is confirmed to be
feasible, and update $UB$ to be $1$ less than the smallest value
confirmed to be infeasible.

\paragraph{Best Fit Decreasing\parperiod} If the values $LB$ and $UB$ are still
unequal, we employ a standard offline bin packing algorithm called
\bfd. \bfd takes items from $\items$ and first sorts them in decreasing
order of their sizes. After that it considers each item one by one in
this order, packing it into a bin where it ``fits best'' -- where it
minimizes the empty space of a bin. We can also interpret it as first
sorting the items in decreasing order and then applying the algorithm
\obf defined above.

As for its complexity, \bfd takes in the worst case $O(m \cdot \items)$
time. It does not need to sort items in $\items$, as the internal
representation of $\items$ keeps the items sorted.

As with \obf, the lower bound $LB$ will updated to the maximum empty
space over all $m$ bins, after \bfd has ended packing. Such an item
can always be sent without invalidating the \binstretching guarantee.

\subsubsection{Procedure $\DynprogMax$}\label{sec:4:dynprogmax}

Procedure \DynprogMax is a sparse modification of the standard dynamic
programming algorithm for \textsc{Knapsack}. Given a multiset $\items,
|\items| = n$ on input, our task is to find the largest item $y$
which can be packed together with $\items$ into $m$ bins (knapsacks) of
capacity $g$ each. 

Instead of initializing the entire DP table, our sparse
approach uses a queue-based algorithm that generates a queue $Q_i$ of
all valid $m$-tuples that can arise by packing the first $i$ items. We
do not need to remember where the items are packed, only the sorted
loads of the bins represented by the $m$-tuple.

To generate a queue $Q_{i+1}$, we initialize it to be an empty queue.
Next, we traverse the old queue $Q_i$ and add the new item
$\items[i+1]$ to all bins as long as it fits, creating up to $m$ new
tuples that need to be added to $Q_{i+1}$.

Unsurprisingly, we wish to make sure that we do not add the same tuple
several times during one step. We can use an auxiliary $\{0,1\}$ array
for this purpose, but we have ultimately settled on a hash-based
approach.

We use a small array $A$ of $64$-bit integers (of approximately $2^{10} -
2^{13}$ elements). When considering a tuple $t'$ that arises from
adding $i$ to one of the bins in the tuple $t$, we first compute the
hash $h(t')$ of the tuple $t'$. Since we use Zobrist hashing (see
Section~\ref{sec:4:caching}), this operation takes only constant time.

Next, we consider adding $t'$ to the queue $Q_{i+1}$. We use the first
$10-13$ bits of $h(t')$ (let $f$ denote their value) and add $t'$ to
$Q_{i+1}$ when $A[f] \neq h(t')$ -- in other words, when the small
array $A$ contains something other than the hash of $t'$ at the
position $f$. We update $A[f]$ to contain $h(t')$ and continue.

While our hashing technique clearly can lead to duplicate entries in
the queue, note that this does not hurt the correctness of our
algorithm, only its running time in the worst case.

We continue adding new items to the tuples until we do $n$ steps and
all items are packed. In the final pass of the queue, we look at the
empty space $e$ in the least-loaded bin. The output of $\DynprogMax$
and the value of $y$ is the maximum value of $e$ over all tuples in
the final pass.

Ignoring the collisions of the hashing scheme (which can happen but
will not play a big role if we compute the expected running time based
on our randomized hashing function), the time complexity of the
procedure \MaxFeas is quite high in the worst case: $\bigO(|\items|\cdot
g^m)$.

Nonetheless, we are convinced that our approach is much faster than
implementing \MaxFeas using integer linear programming or using a CSP
solver (which has been done in \cite{gabay2017improved}) and contributes
to the fact that we can solve much larger instances.
\subsection{Caching}\label{sec:4:caching}

Our minimax algorithm employs extensive use of caching. We cache
solutions of the dynamic programming procedure \MaxFeas as well as any
evaluated bin configuration $C$ (as a hash) with its value.

\paragraph{Hash table properties\parperiod} We store a large hash table of bin
configurations with a 64-bit integer hash. The hash table is addressed
by a prefix of the hash, usually between $20-30$ bits (depending on
the computer used).

We solve collisions by a simple linear probing scheme of a fixed
length (say $4$). In it, when a value needs to be inserted to an
occupied position, we check the following $4$ slots for an empty space
and we insert the value there, should we find it. If all $4$ slots are
occupied, we replace one value at random.

\paragraph{Hash function\parperiod} Our hash function is based on Zobrist
hashing \cite{zobrist}, which we now describe.

For each bin configuration, we count occurrences of items, creating
pairs $(i,f)$ belonging to $\{1,\ldots,g\} \times \{0,1\ldots,m\cdot
g\}$, where $i$ is the item type and $f$ its frequency (the number of
items of this size packed in all $m$ bins).

As for the loads of the $m$ bins, we maintain that they are sorted in
descending order. We also think of them as ordered pairs $(j,g)$, with
$j$ being the position of the bin in the ordering (e.g. $1$ --
largest, $m$ -- smallest) and $g$ the actual value of the load.

For example, we can think of bin configuration
$((3,3,2),\{1,1,1,2,3\})$ as a set of load pairs $(1,3)$, $(2,3)$,
$(3,2)$ along with pairs for items: $(1,3)$, $(2,1)$, $(3,1)$,
$(4,0)$, $(5,0)$ and so on.

At the start of our program, we associate a $64$-bit number with each
pair $(i,f)$. We also associate a $64$-bit number for each possible
load of one bin. These two sets of numbers are stored as a matrix of
size $g \times (m\cdot g)$ and a matrix of size $(t-1) \times m$.

The Zobrist hash function is then simply a \texttt{XOR} of all
associated numbers for a particular bin configuration.

The main advantage of this approach is fast computation of new hash
values.  Suppose that we have a bin configuration $m$ with hash
$H$. After one round of the player \adversary and one round of the
player \algo, a new bin configuration $B'$ is formed, with one new
item placed.

Calculating the hash $H'$ of $B'$ can be done in time $\bigO(m)$,
provided we remember the hash $H$; the new hash is calculated by
applying XOR to $H$, the new associated values, and the previous
associated values which have changed.

\paragraph{Caching of the procedure \MaxFeas\parperiod} We use essentially the
same approach for caching results in the procedure \MaxFeas, except
only the $m$-tuple of loads needs to be hashed.

We also remark upon the values being cached in the procedure \MaxFeas.
At first glance, it seems that it might be best to store the value of
$y$ with each input multiset $\items$. However, this is a very bad
idea, as we would lose upon a lot of symmetry.

Indeed, if we set $i$ to be any item from the list $\items$, we would
lose out on the fact that we know a lower bound on the largest value
that can be sent for a multiset $\items \setminus \{i\} \cup \{y\}$ --
namely $s(i)$, the value we know is compatible.

Instead, it is much better to cache binary feasibilities or
infeasibilities for a specific multiset $\items$. We use these results
to improve the values of $LB$ and $UB$ for other calls of procedure
\MaxFeas.
\subsection{Tree pruning}\label{sec:4:pruning}

Alongside the extensive caching described in Subsection
\ref{sec:4:caching}, we also prune some bin configurations where it is
possible to prove that a simple online algorithm is able to finalize
the packing. Such a bin configuration is then clearly won for player
\algo, as it can follow the output of the online algorithm.

\subsubsection{Algorithmic pruning}\label{sec:4:gs}

Recall that in the game $\Game(m,t,g)$, the player \algo is trying to
pack all items into $m$ bins with load at most $t-1$. If the search
algorithm can quickly deduce that a bin configuration leads to a
successful packing, we can immediately evaluate the configuration as
winning for the player \algo and thus prune the tree.

We can lift several such winning tests -- so-called \emph{good
situations} for the player \algoplayer{} -- from the algorithmic results
of Böhm, Sgall, van Stee and Veselý \cite{bohm2017LB}. However, since
the number of bins $m$ rises from $3$ in \cite{bohm2017LB} up to $8$,
the situations can not always be directly generalized.

We now state the new situations that we have generalized from
\cite{bohm2017LB} for $\Game(m,t,g)$ with $m\ge 4$. For $m=3$, we use
the good situations directly from~\cite{bohm2017LB}.

For the following, we set $\alpha$ to be the extra space that the
player \algoplayer can use without losing, namely $\alpha = (t-1) -g$.
We also make use of the notation from \Cref{sec:notation}.

\begin{goodsit}\label{lem:gs1generic}
Given a bin configuration $(\items,\loads)$ such that the total load
of all but the last bin is at least $(m-1)\cdot g - \alpha$,
there exists an online algorithm that packs all remaining items into
$m$ bins of capacity $t-1$.
\end{goodsit}

\begin{proof}
If the total amount packed is $(m-1)g - \alpha$, the remaining
volume for the instance is $mg - (m-1)g - \alpha = g + \alpha =
t - 1$, which will always fit on the last bin.
\end{proof}

\begin{goodsit}\label{lem:gs2generic}
Given a bin configuration $(\items,\loads)$ such that there exist
two bins $A,B$ such that:

\begin{enumerate}
\item $s(\loads \setminus \{A,B\}) \ge (m-2)g - 2\alpha - 1 $,
\item there exists a bin $C \in \loads \setminus \{A,B\}$ with load below $\alpha$.
\end{enumerate}

then there exists an online algorithm that packs all remaining items
into $m$ bins of capacity $t-1$.
\end{goodsit}

\begin{proof}
We pack the remaining input first into $A$ until an item cannot fit --
we place that item into $C$, where it always fits. After the item is
packed into $C$, the load of $\loads \setminus \{B\}$ satisfies

\[s(\loads \setminus \{B\} )\ge (m-2)g -
  2\alpha - 1 + (g + \alpha + 1) \ge (m-1)g - \alpha,\]
which means Good Situation~\ref{lem:gs1generic} is reached.
\end{proof}

\begin{goodsit}\label{lem:gs3generic}
  Consider a bin configuration $(\items,\loads)$, and let $B_{m}$ stand for the
  least-loaded bin in $\loads$ and $B_{m-1}$ for the second-least-loaded bin.
  Define the following sizes:

\begin{enumerate}
\item Let $s$ be the sum of loads of all bins excluding the last two.
\item Let $r$ (the \emph{last bin load requirement}) be the smallest
number for which the following holds: If the load of $B_m$ is raised to $r$,
GS\ref{lem:gs1generic} is reached (after reordering the bins).
\item Let $o$ (the \emph{overflow}) be defined as $t-r$.
\end{enumerate}

Then, if $r \le t-1$ and if any bin $A \in \{B_{m-1}, B_m\}$ satisfies

\[ \alpha \ge s(A) \ge (m-1) \cdot g - \alpha - o - s, \]

there exists an online algorithm that packs all remaining items into
$m$ bins of capacity $t-1$.
\end{goodsit}

\begin{proof} We first observe that if the bin $B_{m-1}$ is
raised to $r$, GS\ref{lem:gs1generic} is also reached. This follows
from the definition of $r$, which checks for the condition of
GS\ref{lem:gs1generic} that sums the load on all except the least-loaded
bin. The definition of $r$ implies that $B_m$ overtakes $B_{m-1}$, as
otherwise GS\ref{lem:gs1generic} holds immediately. Raising $B_{m-1}$
to $r$ instead (and keeping $B_m$ as the least-loaded bin) leads to
the same calculation, and thus the same conclusion of reaching
GS\ref{lem:gs1generic}.

Let $A$ be the bin satisfying the bound
$\alpha \ge s(A) \ge (m-1)g - \alpha - o - s$ and let $B$
be the other bin from $\{B_{m-1}, B_{m}\}$. 
The algorithm packs greedily into $B$. If $B$ reaches
the threshold load $r$, then GS\ref{lem:gs1generic} is reached. Assuming
the threshold is not reached, there exists a currently unpacked item of size at least $o$
that does not fit into $B$. As $s(A) \le \alpha$, the item $o$ can be packed into $A$.
Summing up loads on $\loads \setminus \{B\}$, we include $s$
(for all bins except $B_{m-1}$ and $B_{m}$), $o$ and the lower
bound on $s(A)$, which sums up to at least
\[s + o + (m-1)\cdot g - \alpha - o - s \ge (m-1)\cdot g - \alpha,\]
which is sufficient for GS\ref{lem:gs1generic}.
\end{proof}

\subsubsection{Adversarial pruning}\label{sec:4:advpruning}

The algorithmic pruning of~\Cref{sec:4:gs}, consisting
of $5$ situations from the literature for $m=3$ and $3$ generic ones
for $m\ge 4$, is a crucial component of our computational approach --
at least according to informal experiments done during our programming
efforts. In contrast, we have
only a few tools to quickly detect that a bin configuration is winning
for the player \adversary. More specifically, we use the
following two criteria:

\paragraph{Large item heuristic\parperiod} Once any bin has load at least $t-g$,
an item of size $g$ packed into that bin would cause it to reach load
$t$, which is a victory for the player \adversary. Suppose that the
$k$-th bin reaches load $l \ge t-g$. We compute the size of the
smallest item $i$ such that

\begin{enumerate}
\item $i + l \ge t$;
\item For any bin $B_t$ with $t \in [(k+1), m]$ it holds
that $s(B_t) + 2l \ge t$; in other words, \algo cannot pack
two items of size $l$ into any bin starting from the $(k+1)$-st.
\end{enumerate}

Finally, we check if \adversary can send $m-k+1$ copies of the item of
size $l$. If so, it is a winning bin configuration for this player and
we prune the tree.

Notice that there may be multiple different values of $l$ for one bin
configuration; {for instance, in the setting of $t=19, g=14$, for $5$ bins
with loads $(10,10,10,9,1)$, we should check whether we can send $2$ items of
size $10$ or $3$ items of size $9$.} Therefore, in the implementation,
we compute for each bin its own candidate value of $l$ and then check
whether at least one is feasible using the dynamic programming test
described in Section~\ref{subsec:test}.

\paragraph{Five/nine heuristic\parperiod} We use a specific heuristic for the
case of $t=19, g=14$, as it is a good candidate for a general lower
bound. This heuristic was experimentally observed to slightly compress
the size of the output tree in this setting.

This heuristic comes into play once there is a bin of load at least
$5$ and once all bins are non-empty (even load $1$ is sufficient). The
item sizes $5$ and $9$ are complementary in the sense that one of each
can fit together in the optimal packing of capacity 14, but the two of
them cannot be packed together into a bin that already has load at
least $5$.

A pair of items of size $9$ also cannot fit together into any other
bin -- as all the bins have already load at least $1$.

Finally, if there are many bins of load at least $5$ and the
guarantees allow an input consisting of sufficiently many items of
size $14$, we may again reach a bin of load at least $19$.

We apply this heuristic only when it is true that at all times, $m$
items of size $9$ can arrive on input without breaking the adversarial
guarantee. While the condition is true, all bins must have load strictly below $10$,
or a load of $19$ is reached immediately.

Our heuristic considers repeatedly sending items of size $5$. If at
any point there are only $p$ bins left with load strictly less than
$5$ and at the same time $p+1$ items of size $14$ can arrive on input,
the configuration is winning for the player \adversary. On the other
hand, if at any point there is a bin of load at least $10$ and the
invariant that $m$ items of size $9$ can still arrive holds, we are
also in a winning state for \adversary.

If it is true that by repeatedly sending items of size $5$ we
eventually reach at least one of the aforementioned two situations,
we mark the initial bin configuration as winning for the player
\adversary.

\paragraph{A note on performance\parperiod} While both of our heuristics reduce
the number of tasks in our tree and the number of considered vertices,
we were unable to evaluate them in every single vertex of the game
tree without a performance penalty. Even the large item heuristic,
which can be implemented with just one additional call to the dynamic
programming procedures of Section~\ref{subsec:test} slows the program
down considerably.

This is likely due to the fact that caching outputs of the dynamic
programming calls of Section~\ref{subsec:test} lead to some vertices
that do not need to call any dynamic programming procedure, while with
our heuristics they are forced to call at least one.

\subsection{Monotonicity}\label{sec:monotonicity}

One of the new heuristics that enables us to go from a lower bound of
$19/14$ on $5$ bins to $8$ bins is iterating on lower bounds by
monotonicity. We define it as follows:

\begin{definition}
A winning strategy for \adversary has \emph{monotonicity} $k$ if it is true
that for any two items $e_i,e_{i+1}$ such that $e_{i+1}$ is sent immediately after
$e_i$, we have $e_{i+1} \ge e_i - k$.
\end{definition}

Using this concept, we can iterate over $k$ from $0$ (non-decreasing
instances) to $g-1$ (full generality) to find the smallest value of
monotonicity which leads to a lower bound, if any.

A potential downside of iterating over monotonicity is that it can
introduce an $g$-fold increase in elapsed time in the case that no
lower bound exists. Additionally, it is quite likely that monotonicity
becomes less useful as the value of $g$ increases, as the item of
relative size $1$ gets smaller and smaller.

Still, solving decision trees of low monotonicity is much faster than
solving the full tree, and we have empirically observed that lower
bounds of lower monotonicity are fairly common; see
Tables~\ref{tab:results3} and \ref{tab:resultsmulti} for our empirical
results.

\paragraph{Monotonicity caveat\parperiod} It is important to remark that when
looking for a lower bound for a specific monotonicity value, it is \emph{not true anymore} that a bin configuration is sufficient
to describe one state of
the bin stretching game. To see this, consider monotonicity $1$. If
the first three input items are $1,2,3$, the next item needs to be of
size $2$ or larger. However, if the three input items are $1,3,2$
(which is permissible for monotonicity $1$), the next item on input
can be of size $1$ and above. This means that the two states are not
equivalent, even though their bin configuration is the same.

To remedy this, we internally extend the definition of the bin
configuration by also marking which item arrived last in the input
sequence, which is sufficient for a fixed value of the monotonicity. %
\subsection{Parallelization}\label{subsec:para}

Up until now, we have described a single-threaded minimax algorithm
with caching and pruning. To get the computing power necessary for
results above $5$ bins, we have implemented the minimax search as a
parallel program for a computer cluster. We now describe the
particulars of this implementation.

\paragraph{Tasks\parperiod} Our evaluation of the game tree proceeds in the
following way: first, we start evaluating the game tree on the main
computer (which we internally call \emph{queen}) until a vertex
corresponding to \adversary's next move meets a certain threshold (for
instance, sufficient depth). After that, we designate this adversarial
vertex as a \emph{task}.

Alongside the queen, we have processes whose job is to evaluate the
tasks -- we call them the \emph{workers}. Workers which run on the
same machine will have a common cache that they access via atomic
primitives in order to maintain consistency. Workers on separate
machines do not share information.

Due to the mixed environment of standard Unix threads and MPI
processes, we also have a single \emph{overseer} per each physical
machine. This overseer handles the MPI communication as well as
spawning the individual worker threads.

The tasks are all generated in advance by the queen. After that, their
bin configurations are synchronized with all overseers running. The
queen then assigns tasks to overseers online, namely by assigning a
batch of 250-500 tasks to an overseer. The overseer reports each value
of a finished task immediately to the queen. When an overseer is
finished processing a batch, it requests and receives a new one.

We have selected this communication strategy for two reasons:

\begin{enumerate}
\item To minimize congestion in the processing phase through the fact
that the bin configurations are synchronized beforehand and only
identifiers are shared in the online assignment phase.
\item To allow the queen to evaluate and prune unfinished tasks and
therefore avoid some unnecessary processing by the workers.
\end{enumerate}

\paragraph{Task selection\parperiod} As mentioned above, an important decision
to be made by the lower bound algorithm designer is where to split a
vertex of the game tree into a task and send it to be processed in the
parallel environment.

Based on our experiments, it seems that maintaining a right balance of
the number of tasks as well as their running time is crucial to good
performance. When the tasks are too shallow, the performance of the
algorithm is dominated by the elapsed time of the most difficult task
in the list, which diminishes the gains coming from the parallel
implementation.

On the other hand, if there are millions of tasks, the algorithms will
still work correctly but we might lose performance from diminishing
advantages of individual caching as well as due to pruning happening
later in the process.

At the outset of our computational efforts, we have only used
\emph{task depth} as the principal guideline -- when $k$ items arrived
on input (with $k$ usually in the range of $\{4,5,6\}$), we mark the
bin configuration as a new task.

Experimenting with running time has shown us that the presence
of larger items speeds up the overall evaluation of the lower bound. One
possible reason may be that large items come with a lower bound on the
upcoming item size, provided monotonicity is set to be small.

Therefore, we have ultimately settled on a mixed task threshold
function which takes into account both the task depth $k$ and also the
\emph{task load} $l$, which is the sum of sizes of all items arrived
so far in the instance. We split off a task when its task load is
above $l$, and failing that when its task depth is below
$k$. After some practical experiments, we have settled on
setting $k \in \{5,6,7\}$ and $l$ to be around $20-40\%$ of the
optimal bin capacity $g$.

\paragraph{Initial strategy\parperiod} Our implementation also allows
us to pre-select some initial strategy for the player \adversary in
advance. This way we can use our (so far limited) intuitive
understanding of what is a good initial move and decrease the time
needed to evaluate the whole tree.

A particularly good strategy for the lower bound of $19/14$ ($t=19,
g=14$) seems to be sending an item of size $5$ as the first item,
followed by several (5 in the case of $m=8$) items of size $1$. This
adversarial strategy leads to a lower bound instance for $6,7$ and $8$
bins.

We have therefore implemented a way to pre-select items to be sent in
the first few rounds of the game. Given such a list of items, we
compute all possible moves of the player \algo and create a queue of
bin configurations that we each evaluate sequentially.

The fact that already this linear, non-adaptive strategy of sending
$5,1,1,\ldots$ is enough to get a lower bound of $19/14$ for $8$
bins was a pleasant surprise to us. We believe this fact is due to the
size of the sequence being already non-trivial (the item $5$ alone
occupies slightly more than $25\%$ of one stretched bin).

A natural extension is to allow the user to input a partial game tree
(an adaptive strategy for the player \adversary) and have the
algorithm evaluate it sequentially; this can be easily added to our
implementation once we learn more about which items should be the
among the first to send.

\paragraph{Technology\parperiod} We have settled on using a combination of
OpenMPI \cite{openmpi} and the standard thread library as provided by the
C++ programming language. In our setting, OpenMPI is used to provide
inter-computer communication API for sending and receiving tasks as
described above. We employ the standard Unix threads to spawn the
worker processes themselves; this way they can easily share one large
cache for evaluated bin configurations.

We have originally considered using only OpenMPI processes for both
inter-computer communication as well as memory sharing on one physical
computer; this functionality is present in the latest version of the
MPI standard, MPI-3.0. However, after implementing the shared memory
functionality, we have noticed some slowdown of the worker processes
when the shared memory was large (more than 1 gigabyte). This forced
us into the heterogeneous model that we use right now.

\subsection{Results}\label{sec:results}

Tables~\ref{tab:results3} and \ref{tab:resultsmulti} summarize our results; we
include previous results for completeness. Note that there may be a lower bound
of size say $\frac{41}{30}$ even though none was found when computing $t = 41, g
= 30$; the same bound can be achieved, possibly, by setting $t = 164 , g = 120$,
which is beyond the capabilities of our current program. %
{In addition, it is unreasonable to formally certify the correctness of the
program result when no lower bound is found, so negative outputs should not be
considered as definitive results. We nevertheless also report in
Table~\ref{tab:results3} some candidate fractions for which the program
terminates without finding a lower bound. This allows some insights on values
leading to negative results and the computing time required to explore the whole
solution space. We can see for instance that the fractions $\frac{30}{22}$ and
$\frac{56}{41}$ did not lead to a lower bound although the decimal value of
$\frac{112}{82}$ is not smaller. 
}

\begin{table}[H]
\begin{center}
\begin{tabular}{ llllll }
 & & & &  \multicolumn{2}{c}{\textit{Elapsed time}}  \\
\textit{Fraction} & \textit{Decimal} & \textit{L. b.} & \textit{Mon.} & \textit{Linear} & \textit{Parallel}\\

\hline
$19/14$ &  $1.3571$ & Yes & 0 & 2s. & \\
$22/16$ & $1.375$ & No & & 2s. & \\
$26/19$ & $1.3684$ & No & & 3s. & \\
$30/22$ & $1.\overline{36}$ & No & & 6s. & \\
$33/24$ & $1.375$ & No & & 5s. & \\
$34/25$ & $1.36$ & Yes & 1 & 15s. & \\
$41/30$ & $1.3\overline{6}$ & No & & & \\
$45/33$ & $1.\overline{36}$ & Yes & 1 & 1min. 48s. & \\
$55/40$ & $1.375$ & No & & 3min. 6s. & \\
$56/41$ & $1.3659$ & No & & 30min. & 7s. \\
\hline
$82/60$ & $1.3\overline{6}$ & No & & & 21 m. 49s. \\
$86/63$ & $1.36507$ & \textbf{Yes} & 6 & & 29s. \\
$112/82$ & $1.3659$ & \textbf{Yes} & 8 & & 3h. 21m. 31s.\\
\end{tabular}
\end{center}
\caption[LoF entry]{The results and performance of our linear and
parallel computations for \binstretching with three bins. The results
above the horizontal line were previously shown in \cite{gabay2017improved} and
\cite{bohm2016}. The column \textit{L. b.} indicates whether a
lower bound was found when starting with the given stretching factor
$t/g$ as seen in column \textit{Fraction}.
The column \textit{Mon.} shows the lowest monotonicity that our
program needs to find a lower bound. In the case of negative results,
time measurements were done only using full generality, i.e. with
monotonicity $g-1$. 
The linear results were computed on a server with an AMD Opteron 6134
CPU and 64496 MB RAM. The size of the hash table was set to $2^{25}$.
The parallel results were computed using OpenMPI on a heterogeneous
cluster with $109$ worker processes running.
The output of the program was not generated during the
time measurements.}
\label{tab:results3}
\end{table}

\begin{table}[H]
\begin{center}
\begin{tabular}{lllllll}
& & & & & \multicolumn{2}{c}{\textit{Elapsed time}}  \\
\textit{Bins} & \textit{Fraction} & \textit{Decimal} & \textit{L. b.} & \textit{Mon. (5)} & \textit{Linear} & \textit{Parallel (5)}\\
$4$  & $19/14$ &  $1.3571$ & Yes & & & 18s.  \\
$5$  & $19/14$ &  $1.3571$ & Yes & 2 (1) & & 10s. \\
\hline
$4$  & $30/22$ & $1.\overline{36}$ & No   & & & 19s. \\
$4$  & $34/25$ &  $1.36$   & No           & & & 48s.  \\
$4$ & $45/33$ & $1.\overline{36}$ & No & & & 1h. 1m. 40s. $\dagger$ \\
$6$  & $19/14$ &  $1.3571$ & \textbf{Yes} & 0 (0) & & 11s. \\
$7$  & $19/14$ &  $1.3571$ & \textbf{Yes} & 1 (0) & & 2m. 13s. (16s.) \\
$8$  & $19/14$ &  $1.3571$ & \textbf{Yes} & Unk. (1) & & (1h. 14s.)  \\
\end{tabular}
\end{center}
\caption[LoF entry]{The results produced by our minimax algorithm for more than $3$ bins.
Tested on the same machine and with the same parameters
as in Table~\ref{tab:results3}, both for linear and parallel
computations. The result $\dagger$ was computed subsequently
in a parallel environment with 64 threads and 512 MB of shared cache.

In columns \textit{Mon.} and \textit{Parallel}, we list in brackets
monotonicity and elapsed time of computation for an input having an
item of size 5 at the start. Monotonicity is measured only starting
with the second item.}
\label{tab:resultsmulti}
\end{table}

\section{Certification}
\label{sec:coqform}

We describe in this section how we certify the results obtained
using the computer search of \Cref{sec:search} via the Coq proof
assistant system.  The aim of this section, as explained in the
introduction, is to formalize the \emph{lower bound property} -- a
predicate being true if there exists a valid lower bound -- in Coq,
prove that this property as we define it matches the intended meaning,
and comment on the technical challenges that were needed to be
overcome.

We first describe the Coq formalization of the problem previously defined in \Cref{sec:def}.
In \Cref{sec:coqprelim}, we define the relevant types and preliminary
functions. In \Cref{sec:coqdef}, we describe the core of our
formalization. Specifically, we first define the function updating the
bin configuration after the addition of a given item in some bin. We
then define an inductive predicate that recognizes a winning strategy
for \adveplayer, given a bin configuration. We finally use this
predicate to define the main predicate \lbcoq. We then show in
\Cref{sec:correct} that this formalization is correct: if the Coq
predicate \lbcoq is true, then the property \lb is true. Indeed, the
goal of the Coq script is to prove that \lbcoq is true for given
values of $m$, $t$, $g$ defining the game $\Game(m,t,g)$. In
\Cref{sec:correct}, we therefore show that this result actually
implies a lower bound on this game, {\it i.e.,} implies \lb.
Finally, in \Cref{sec:verif}, we present the results obtained on the
files generated by the program described in \Cref{sec:search} and
detail some features that had to be implemented in order to handle the
large file sizes involved.

The code is available online at \cite{GithubCoq}, and also contains a program
which translates an adversary strategy expressed using the widespread
GraphViz format into a file which can be directly processed by our Coq
script. Then, future lower bounds can be easily certified using the
same script.

The first part of this section can serve as an introduction to
formalization of an online lower bound into the language of Coq, and
we hope it will be interesting even to readers with primarily
theoretical focus.

\subsection{Preliminaries}
\label{sec:coqprelim}

We start with defining variables $m, t, g$, which correspond to
the parameters of Bin Stretching from \Cref{sec:search} -- $m$ is the
number of bins, $t$ is the target capacity (how much the player \algo
is allowed to pack), and $g$ is the guarantee capacity (how much the
player \adversary is allowed to pack). We also require that $m$ is strictly positive,
although for any intended use all three variables will be positive.

\begin{center}
\begin{coq}
Variables m t g : nat.
Hypothesis Posm : m > 0.
\end{coq}
\end{center}

In order to distinguish some objects in our properties by a
specific name, much like in a programming language, we define a few
key data types. The type \verb|BinExtended| represents the list of items present in a
given bin, and is then implemented as a list of integers. The
type \verb|BinLoads| represents the current load of all bins, and is
then also implemented as a list of integers, one per bin. The
\verb|BinsExtended| type corresponds to a bin configuration in its extended
representation (see \Cref{dfn:notation}), and is internally
represented as a list of types \verb|BinExtended|, one per bin.

\begin{center}
  \begin{coq}
Definition BinExtended  := list nat.
Definition BinLoads   := list nat.
Definition BinsExtended := list BinExtended.\end{coq}
\end{center}
  
For any list of integers, the property \verb|Iszero| is true if and
only if the list contains only zeros and at most $m$ items. It
represents the starting loads of the bins, where some bins may be
omitted.

The simplest recursive functions in Coq can be defined using
the \verb|Fixpoint| keyword, which can be used only when the
recursion is applied to a simple inductive object, such as a list.
The following basic functions are defined in this manner. The function
\verb|BinSum| returns the load of a bin, given the list of items
present in this bin. The function \verb|MaxBinSum| returns the load of
the highest (most-loaded) bin, given the bin configuration, and the
function \verb|MaxBinValue| returns the load of the highest bin given
only the vector of loads. Note that \verb|nil| represents the empty
list and \verb|x::s| represents the list of head \verb|x| and tail
\verb|s|.

\begin{center}
\begin{coq}
Definition Iszero l := (length l <= m) /\ (forall e, In e l -> e = 0).

Fixpoint BinSum (B: BinExtended) := match B with
| nil   => 0
| x ::s => x + BinSum s
end.

Fixpoint MaxBinSum (P: BinsExtended) := match P with
| nil   => 0
| x ::s => max (BinSum x) (MaxBinSum s)
end.

Fixpoint MaxBinValue (\codeloads: BinLoads) := match \codeloads with
| nil   => 0
| x ::s => max x (MaxBinValue s)
end.
\end{coq}
\end{center}

A \verb|Fixpoint| can also be defined over natural numbers,
where the natural numbers themselves -- defined with the successor
function \verb|S k| as in the Peano arithmetic, defined to be equal to $k+1$ -- serve as the inductive
structure. We use this type of recursion to define the function
$\verb|AddToBin|$ that models adding an item to a bin. The function
takes three parameters: $\codeloads$ of type \verb|BinLoads| and two integers $e$
and $b$. This function increases the load of the $b$-th bin by a value
equal to $e$. If $b$ is larger than the length of $\codeloads$, a new item
of value $e$ is appended to $\codeloads$.

\begin{center}
\begin{coq}
Fixpoint AddToBin ($\codeloads$: BinLoads) (e: nat) (b: nat) := match $\codeloads$,b with
| nil  , b     => [e]
| x ::s, 0     => (x+e) ::s
| x ::s, (S k) => x :: (AddToBin s e k)
end.
\end{coq}
\end{center}

\subsection{Defining the main predicates}
\label{sec:coqdef}

We now define a few properties specific to the \binstretching problem.

The predicate \verb|SequencePacked|, given a list of item sizes
(integers) $\codeitems$ and an element $P$ of type \verb|BinsExtended|, is
true if the configuration $P$ uses at least all the items of
the list (sequence) $\codeitems$. It uses two functions which are part of the Coq standard
library. The function \verb|count_occ Nat.eq_dec x y| returns the
number of occurrences of the element $y$ in the list $x$ and the
function \verb|concat| concatenates a list of lists of elements.

\begin{center}
\begin{coq}
Definition SequencePacked ($\codeitems$ : list nat) (P: BinsExtended) := forall e, 
count_occ Nat.eq_dec $\codeitems$  e <= count_occ Nat.eq_dec (concat P) e.
\end{coq}
\end{center}

The predicate \verb|GuaranteePacking|, 
given the same parameters as the predicate
\verb|SequencePacked|, is true if \verb|SequencePacked| is true, the
length of $P$ is equal to $m$ and no bin has load larger than $g$.
Such a packing $P$ is then a certificate that the items described in
$\codeitems$ can be packed in $m$ bins of capacity $g$.

\begin{center}
\begin{coq}
Definition GuaranteePacking ($\codeitems$ : list nat) (P: BinsExtended) :=
SequencePacked $\codeitems$  P /\  length P = m /\ MaxBinSum P <= g. 
\end{coq}
\end{center}

The main predicate used in the formulation is \verb|OnlineInfeasible|,
which is a parametric predicate with three variables: an integer
$\mathtt{X}$, a list $\codeitems$, and $\codeloads$, which is one value of
\verb|BinLoads|. The list $\codeitems$ corresponds to items being sent on input
initially, and the state of bins $\codeloads$ corresponds to one
algorithmic packing of items from $\codeitems$ into $m$ bins. The auxiliary
variable $\mathtt{X}$ is not necessary in the definition, but it allows the Coq
prover to easily assume an induction hypothesis when inductively
proving properties of \verb|OnlineInfeasible|.
  
We employ the Coq syntax to distinguish only two cases when this
predicate is true, naming them \verb|Overflow| and
\verb|Deadend|. In the case of  \verb|Overflow|,
  one bin of $\codeloads$ is loaded to at least the value $t$, and at
  the same time it still holds that there exists an optimal packing
  for the current input list $\codeitems$.

The other case when \verb|OnlineInfeasible| is true,
which we call \verb|Deadend|, occurs when we can recursively
reach the state \verb|OnlineInfeasible| being true by
presenting a new item $e'$ (a positive integer) on input, for all
choices of adding $e'$ into any of the bins. For practical reasons, we
use a non-negative integer $e$ as the variable used for the new item,
and define $e' = e+1$, which can be written as ``\texttt{S e}'' in the Coq
code. This is one of the simplest ways to define a positive (nonzero) integer.
  
The full statement of \verb|OnlineInfeasible| is as follows:

\begin{center}
\begin{coq}
Inductive OnlineInfeasible: nat -> list nat -> BinLoads -> Prop :=
| Overflow X $\codeitems$ $\codeloads$:   t $\le$ MaxBinValue $\codeloads$  -> (exists P, GuaranteePacking $\codeitems$  P) 
	-> OnlineInfeasible X $\codeitems$ $\codeloads$
| Deadend  X $\codeitems$ $\codeloads$:   length $\codeloads$ $\le$ m   
	-> (exists e, forall b, (b < m) 
		-> OnlineInfeasible X ( (S e) ::$\codeitems$) (AddToBin $\codeloads$ (S e) b) )
	-> OnlineInfeasible (S X) $\codeitems$ $\codeloads$.
\end{coq}
\end{center}

The syntax implies the following equivalence.
\begin{align}
\forall~ \texttt{X}, \codeloads, \codeitems \colon\quad & \texttt{{OnlineInfeasible (X+1) $\codeitems$ $\codeloads$}}~\Longleftrightarrow \label{eq:OIX}\\ 
&  \Big( \left(t \leq \texttt{MaxBinValue $\codeloads$} ~\wedge~ \exists \texttt{P}\colon \texttt{GuaranteePacking $\codeitems$ P}\right)\nonumber\\
& \vee~ \big(\texttt{length $\codeloads$} \leq m
~ \wedge~ \exists e'>0, \forall b<m\colon \nonumber\\
&~~~~~~~ \texttt{OnlineInfeasible X ($e'$::$\codeitems$) (AddToBin $\codeloads$ $e'$ b)}
\big)\Big) \nonumber
\end{align}

The final predicate defined is:

\begin{center}
\begin{coq}
Definition LowerBoundCoq := exists s, 
Iszero s  /\  OnlineInfeasible (m*g+2) [ ] s.
\end{coq}
\end{center}

The value $m\cdot g+2$ in the definition of the predicate is there as
a simple upper bound of the number of inductive steps sufficient for
any correct proof (recall that no more than $mg$ items can arrive in
any valid input for \binstretching).

\subsection{Correctness of the Coq formulation}
\label{sec:correct}

The Coq proof assistant can now be used to simplify our proving
efforts, using our computed instances as a proof strategy to verify
that the proposition \texttt{LowerBoundCoq} is true. What remains to
be formally proven outside of the Coq system is to show that the
propositions from \Cref{sec:coqdef} actually match what we actually
wish to compute, namely a lower bound on the bin stretching game.

First, let us restate the winning properties for bin stretching
  from \Cref{sec:def}:

\propertylb*
  
The final piece of the puzzle is to show the following theorem,
which immediately implies the correctness of the Coq formulation
(\Cref{coro:correct}).

\begin{theorem}\label{thm:lastpiece}
	For any $\codeitems, \codeloads, \mathtt{X},$  the proposition
	\textnormal{\texttt{OnlineInfeasible X $\codeitems$ $\codeloads$}} implies $\lb(\codeitems,\codeloads)$.
\end{theorem}

\begin{coro}
	\label{coro:correct}
	\lbcoq implies \lb.
\end{coro}

\begin{proof}[Proof of \Cref{thm:lastpiece}.]

	We prove this result by reverse induction on the sum of the items of $\codeitems$. Let $\codeitems$ and $\codeloads$ be two lists of integers and $\mathtt{X}$ be an integer.
	
	We make use of the decreasing parameter $\mathtt{X}$ and the fact that $e'>0$ in the definition of \oi, {to prove the following property in Coq (validated in supplementary code as Theorem} \verb|OI_length|):
	$$
	\forall \mathtt{X}, \codeitems, \codeloads,~ \mathtt{OnlineInfeasible~ \mathtt{X}~ \codeitems~ \codeloads} ~ \Longrightarrow (\mathtt{BinSum ~\codeitems} \leq mg).
	$$
	
	Therefore, for a list $\codeitems$ with a large sum,  $\oi(\mathtt{X}, \codeitems, \codeloads)$ is false for any value of $\mathtt{X}$ and $\codeloads$, making \Cref{thm:lastpiece} (an implication) true.
	
	\smallskip
	
        Moving on, we fix the value of $\mathtt{X}$.
	Let $L\geq 0$ and suppose by induction that for all $\codeitems$ whose items sum to more than $L$, for all  $\codeloads$, the proposition
	$(\mathtt{OnlineInfeasible~ X ~\codeitems~ \codeloads})$ implies the property $\lb(\codeitems,\codeloads)$.
	
	Consider any $\codeitems$, a list whose items sum to exactly $L$, and any $\codeloads$ such that we have $\oi(\mathtt{X}, \codeitems, \codeloads)$.
	
	We want to show the property $\lb(\codeitems,\codeloads)$. Using \Cref{eq:OIX} and the proposition $\oi(X, \codeitems, \codeloads)$, we have two cases.
	
	First, if $\left(t \leq \verb|MaxBinValue \codeloads| ~\wedge~ \exists P~,~ \texttt{GuaranteePacking $\codeitems$ P}\right)$\linebreak holds, then one bin of \codeloads has load at least $t$ and there exists a packing of the items of $\codeitems$ into $m$ bins with load at most $g$. Therefore,  the property $\lb(\codeitems,\codeloads)$ is true.
	
	Otherwise, there exists a number $e'>0$ such that the following property holds.
	\begin{align}
	&\texttt{length \codeloads} \leq m ~~~ \wedge~ \nonumber\\ 
	&\forall b<m,~ \texttt{OnlineInfeasible (X-1) ($e'$::$\codeitems$) (AddToBin $\codeloads$ $e'$ b)}\label{eq:OIx-1}
	\end{align}

	Note that the value of $X-1$ can be replaced by $\mathtt{X}$ as the following is true (and {proved in Coq as Theorem} \verb|OI_Succ|):
        If the predicate $\texttt{OnlineInfeasible}$ is true when parameterized by $X-1$, then it is also true when parameterized by $\mathtt{X}$.
	
	Consider any possible move $b$ for \algoplayer after \adveplayer played $e'$.	
	Using \Cref{eq:OIx-1} and the induction hypothesis, we know that the property $\lb(e'::\codeitems, (\texttt{AddToBin \codeloads\ } e'\ b))$ holds. So, after \algoplayer played $b$, \adveplayer has a winning strategy.
	
	As this is true for all possible moves $b$ of \algoplayer, we have the property $\lb(\codeitems,\codeloads)$, which completes the proof.
\end{proof}

\subsection{Verification of a winning strategy for \adveplayer}
\label{sec:verif}

We now detail how we used the results obtained in \Cref{sec:search} in
order to prove the property \lbcoq for a given game $\Game(m,t,g)$. We
rely on a file, computed by the aforementioned program, which
describes a winning strategy for \adveplayer: which moves he makes
after each possible move of \algoplayer, as well as the packing
solutions on winning states. The format is based on the tree structure
illustrated in \Cref{fig:lb43}, with several improvements described
below. In order to verify that this file is a correct representation
of a lower bound, we implement in Coq a function that performs
multiple checks, which we call \verb|Check| in this section. In
essence, this function is analogous to the verifier program discussed
in \Cref{sec:intro}. The crucial difference is that \verb|Check| is
certified: a theorem, proven in Coq, states that, if \verb|Check|
returns \verb|true|, then the predicate \lbcoq is valid for the game
$\Game(m,t,g)$. Then, by \Cref{coro:correct}, \lb is also valid for
the game $\Game(m,t,g)$. 

Although we do not detail the complete Coq script here, which exceeds
2000 lines~\cite{GithubCoq}, we would like to emphasize that the format used to store
the lower bound as well as the function \verb|Check| that verifies it
are not implemented in a naive manner because of the file sizes
involved. The features implemented, which therefore complicate the Coq
script proving the correctness of \verb|Check|, include the following.

\paragraph{DAG encoding\parperiod}

The naive tree decomposition of a winning strategy for \adveplayer
details every decision that has to be made, but may contain a large
number of duplicate subtrees. Indeed, several nodes of the tree
correspond to the same list of items and loads of bins (up to
irrelevant permutations). We therefore use a DAG structure to store
these duplicates. As constant-access data structures are not available
in Coq, we use a single list of trees $R$ to denote all the existing
duplicate subtrees. When examining the possible outcomes from a node
according to the decision of \algoplayer, there are then three
possibilities: it corresponds to a direct child of this node (if such
a subtree is unique), it corresponds to a tree in the list $R$, or one
bin exceeds the target load. Note that trees of the list $R$ can
themselves refer to subsequent trees of the same list, and we are then
able to prove our results by induction. It remains to implement a fast
way to check that an item is present in the next part of the list
$R$. We use for this purpose an AVL tree dictionary indexed by a pair
of lists describing the current items and bin loads. To assess the
importance of removing these duplicates, notice on \Cref{tbl:LBtime}
that it decreased the largest graph size by three orders of magnitude.

\paragraph{Last layer compression\parperiod} 

Often, the last items that are sent by the adversary are independent
from the decisions made by \algoplayer. However, they can represent a
large portion of the nodes in the normal DAG representation. Hence, we
store such a situation only as a single node with a list of upcoming
items, instead of the full tree. This corresponds directly to storing
only one node when the \textit{large item heuristic} of
\Cref{sec:4:advpruning} is successful. The number of nodes obtained in
such a compressed DAG (cDAG) is represented in \Cref{tbl:LBtime}, and
leads to a decrease by a factor of 5.

\paragraph{Binary integers\parperiod} 

Coq proofs often rely on the Peano arithmetic, where a natural integer is
represented in a unary way by being either 0 or the successor of an
integer. In order to decrease the time and resources required to prove our
results, we perform the computations using a binary integer
representation. We have therefore implemented two analogous functions,
which we can name \verb|Check_binary| and \verb|Check_unary|, working
respectively on binary and unary integers. We prove that these
functions give the same result and that if \verb|Check_unary| returns
true, then \lbcoq is valid. Therefore, we can run the function
\verb|Check_binary| while using unary arithmetic in most of our
proofs.

\medskip

\begin{table*}[tbh]
	\begin{tabular}{ccccccccc} \toprule
		Value of $m$   & $3$ & $4$ & $5$ & $6$ & $7$ & $8$ \\ \midrule
		Lower bound    & 112/82 & 19/14 & 19/14 & 19/14 & 19/14 & 19/14 \\
		Tree nodes    & $186$k & 433 & 3908 & $3.8$M & $231$M & $ 2.5$G\\		
		DAG nodes      & $103$k & 236 & 1271 & $38$k & $186$k & $1.6$M \\		
		cDAG nodes     & $37$k & 102 & 408 & 7k & 61k & 598k \\		
		Time           & 38s & 1s & 2s & 12s & 4m30 & 2h & \\
		\bottomrule
	\end{tabular}
	\centering
	\caption{Size of the uncompressed and compressed DAGs and (approximate) time needed to load the trees and certify each lower bound. The running times were computed on a machine with the Intel Core i5-6600 CPU and 32 GB of RAM.}
	\label{tbl:LBtime}
\end{table*}

With these features implemented, we have been able to certify all the
lower bound results previously published and presented in this paper.
The amount of time necessary to run each Coq script is reported in
\Cref{tbl:LBtime}.

\bibliography{mybibfile}

\end{document}